\documentclass[%
 aip,
 jap,
 amsmath,amssymb,
 reprint,%
]{revtex4-1}

\usepackage{graphicx}
\usepackage{dcolumn}
\usepackage{bm}

\usepackage[utf8]{inputenc}
\usepackage[T1]{fontenc}
\usepackage{mathptmx}

\begin{document}

\preprint{AIP/123-QED}

\title{Charge coloration dynamics of electrochromic amorphous tungsten oxide studied by simultaneous electrochemical and color impedance measurements}

\author{Edgar A. Rojas-Gonz{\'a}lez}
 \email{edgar.rojas@angstrom.uu.se.}

\author{Gunnar A. Niklasson}%
\affiliation{ 
Department of Materials Science and Engineering, The {\AA}ngstr{\"o}m  Laboratory, Uppsala University, P.O. Box 35, SE-751 03 Uppsala, Sweden
}%

\date{\today}

\begin{abstract}
The coloration mechanisms in electrochromic (EC) systems can be probed by comparing the dynamics of the electrical and optical responses. In this paper, the linear frequency-dependent electrical and optical responses of an amorphous tungsten oxide thin film were measured simultaneously by a combination of two techniques{\textemdash}that is, electrochemical impedance spectroscopy (EIS) and the so-called color impedance spectroscopy (CIS). This was done at different bias potentials, which can be associated with different intercalation levels. Equivalent circuit fitting to the EIS spectra was used to extract the Faradaic components participating in the total impedance response. The latter were assigned to an intermediate adsorption step before the intercalation and to the diffusion of the electron-ion couple in the film. A quantity denoted complex optical capacitance is compared to the complex electrical capacitance{\textemdash}in particular, their expressions are related to the Faradaic processes. The coloration at low intercalation levels followed both the adsorption and diffusion phenomena. Conversely, the diffusion contribution was dominant at high intercalation levels and the adsorption one seemed to be negligible in this case. For perfectly synchronized electrical and optical responses, their complex spectra are expected to differ only by a multiplying factor. This was the case at low intercalation levels, apart from small deviations at high frequencies. A clear departure from this behavior was observed as the intercalation level increased. A combination of frequency-dependent techniques, as presented in this work, can help in the understanding of the dynamics of the coloration mechanisms in EC materials at various conditions{\textemdash}for example, at different intercalation levels and optical wavelengths.
\end{abstract}

\maketitle

%
\section{\label{sec1:level1}Introduction}

Electrochromic (EC) materials possess the ability to adjust their optical properties by means of an external electrical stimulus.\cite{Granqvist1995} This feature makes them of great interest for relevant technological applications; for example, energy-efficient smart windows.\cite{Granqvist2019} Here, EC inorganic oxide thin films are sandwiched between a transparent conductor and an electrolyte containing small ions. By controlling an external voltage, ions can be reversibly intercalated into the EC coating. The latter results in an optical modulation produced by the corresponding electrons that are inserted into the EC film (from the outer circuit) due to charge neutrality requirements.\cite{Granqvist1995} Tungsten oxide ($\mathrm{WO}_3$) is a widely studied and used EC material.\cite{Niklasson2007} The term tungsten oxide, $\mathrm{WO}_3$, is used here referring to the general case without specifying its form (amorphous or crystalline). In the case of amorphous tungsten oxide ($\textit{a}\mathrm{WO}_3$), its optical absorption{\textemdash}related to the extra inserted electrons{\textemdash}is assigned to intervalence charge transfer between localized tungsten sites.\cite{Faughnan1975}

Dynamic EC processes involve several steps taking place in different sequences. For example, charge accumulation and intercalation at the electrolyte-EC film interface, ion and electron diffusion within the EC material, and other effects due to the outer circuit.\cite{Crandall1976,Ho1980} In this regard, linear frequency-dependent methods, like electrochemical impedance spectroscopy (EIS), are particularly useful for decoupling the effects of the different contributions by identifying their characteristic frequency response.\cite{Bonanos2018}

EIS provides the relation between the oscillating voltage and current in the linear regime of an EC system. In this case, it is natural to include the corresponding optical response into the frequency-dependent analysis. This is usually referred to as color impedance spectroscopy\cite{Amemiya1993} (CIS) and was first introduced for inspecting the kinetics of adsorption processes at the electrolyte-electrode interface.\cite{Adzic1973} Thereafter, it has been employed in the study of diverse systems, such as polyaniline,\cite{Greef1989,Hutton1989,Kalaji1991} nickel oxide,\cite{SusanaInesCordobaTorresi1990} polypyrrole,\cite{Amemiya1993,Amemiya1993b,Amemiya1993a,Amemiya1994} tungsten oxide,\cite{Gabrielli1990,Gabrielli1994,Kim1997a,Kim1997,Garcia-Belmonte2004,Bueno2008,Rojas-Gonzalez2019,RojasGonzalez2020} hemin and nile blue A,\cite{Feng1995} cytochrome c proteins,\cite{Ruzgas1998,Araci2008,Han2013,Han2014,Xue2017} Alexa 488 fluorochromes,\cite{Li2000} Prussian blue,\cite{Araci2006,Agrisuelas2009a,Agrisuelas2012,Agrisuelas2015} poly(3,4-ethylenedioxythiophene) and poly(3,4-ethylenedioxythiophene methanol),\cite{Doherty2006} niobium oxide,\cite{Bueno2008} pyridine-capped $\mathrm{CdSe}$ nanocrystals,\cite{Araci2010} and graphite.\cite{Manka2015} Here, CIS is circumscribed to optical impedance measurements{\textemdash}performed over a range of modulating frequencies{\textemdash}that probe the dispersion of the relevant complex impedance quantities. Hence, this definition does not include similar approaches using only one excitation frequency, presented elsewhere.\cite{Seraphin1965,Cardona1967,Venkateswara1986,Sagara1991}

In the context of small sinusoidal perturbations, the complex capacitance provides the amplitude and phase relation between the oscillating charge and voltage signals. For EC systems, there is a close connection between the state of charge and optical absorption.\cite{Granqvist2019} Thus, making an analogy between these quantities, it is possible to define a complex optical capacitance which relates the oscillating absorption response to the oscillating voltage signal{\textemdash}this concept was introduced by Kalaji and Peter in Ref. [\onlinecite{Kalaji1991}]. Provided that the electrochromic effect is due to a single redox reaction involving only two species,\cite{Kalaji1991} the complex optical capacitance should be proportional to the part of the complex capacitance that is related to the Faradaic charge involved in the coloration. However, this simple proportionality relation breaks down if three or more species (resulting in multiple redox reactions) participate in the absorption processes.\cite{Kalaji1991} Regarding $\mathrm{WO}_3$ studies, interesting works combining EIS and CIS can be found in the literature. In this case, discrepancies between the complex optical and electrical capacitances have been attributed to diverse origins. Some examples include leakage currents,\cite{Kim1997,Garcia-Belmonte2004} the existence of different intercalation sites in the $\mathrm{WO}_3$ host associated with dissimilar absorption strengths,\cite{Gabrielli1990,Gabrielli1994} a coloring ionic trapping process with an intermediate ion diffusion step before reaching a trapping state where the coloration can take place,\cite{Garcia-Belmonte2004,Bueno2008} and non-Faradaic high-frequency capacitive effects\cite{Kim1997} which may originate from $\mathrm{ClO}_4^-$ intake by the $\mathrm{WO}_3$ film\cite{Gabrielli1994} (when a $\mathrm{LiClO}_4$-propylene carbonate electrolyte is used). As a first approximation, the EC phenomenon can be ascribed to a single redox reaction related to transitions between two species{\textemdash}that is, $\mathrm{W}^{5+}$ and $\mathrm{W}^{6+}$ sites. However, the influence of $\mathrm{W}^{4+}$ sites on the EC effects of $\mathrm{WO}_3$ has been highlighted by previous studies,\cite{RojasGonzalez2020,Zhang1997,Bueno2004,Darmawi2015} and is more relevant at higher intercalation levels in accordance with an extended site-saturation theory.\cite{RojasGonzalez2020,Berggren2007} Hence, as the intercalation level increases, it is expected to observe a transition between a coloration process dominated by a single redox reaction{\textemdash}involving mainly $\mathrm{W}^{5+}$ and $\mathrm{W}^{6+}$ sites{\textemdash}toward a more intricate situation due to the increasing relevance of the $\mathrm{W}^{4+}$ sites. This possible source of discrepancies between the complex Faradaic capacitance and the complex optical capacitance has not been addressed in detail before for $\mathrm{WO}_3$ in general, and its study is important for improving our understanding of EC processes and their frequency response. 

It is worth mentioning that, at potentials below about $2~\mathrm{V}~\mathrm{vs}.~\mathrm{Li}/\mathrm{Li}^+$, irreversible lithium trapping processes\cite{Arvizu2015,Baloukas2017} may influence both the electrical and optical spectra{\textemdash}in fact, CIS measurements on $\textit{a}\mathrm{WO}_3$ have shown deviations from the equilibrium condition when they were performed at bias potentials lower than $2~\mathrm{V}~\mathrm{vs}.~\mathrm{Li}/\mathrm{Li}^+$.\cite{RojasGonzalez2020} These effects could also contribute to a mismatch between the spectra at the mentioned potential range.

In this work, we focus on the specific $\textit{a}\mathrm{WO}_3$ case and show simultaneous EIS and CIS\cite{Rojas-Gonzalez2019} (SECIS) measurements on $\textit{a}\mathrm{WO}_3$ thin films at an optical wavelength of $810~\mathrm{nm}$ and at different bias potentials (or intercalation levels). The EIS spectra are fitted to an equivalent circuit containing impedance elements assigned to anomalous diffusion\cite{Bisquert2001} and an intermediate adsorption step before the intercalation.\cite{Franceschetti1982} The complex capacitance resulting from the Faradaic part of the equivalent circuit is compared, at different intercalation levels, to the complex optical capacitance. In addition, the individual contributions of the diffusion and adsorption processes to the complex Faradaic capacitance are compared to the complex optical capacitance at each intercalation level.

\section{\label{sec2:level1}Experimental setup and procedures}

\subsection{\label{sec2:level2_1}Electrode preparation}

Amorphous tungsten oxide $\textit{a}\mathrm{WO}_3$ thin films were prepared by reactive DC magnetron sputtering in a deposition system which is based on a Balzers UTT 400 unit. The deposition took place from a $99.95\%$-pure $\mathrm{W}$ target (5-cm-diameter disc) simultaneously onto unheated glass substrates pre-coated with $\mathrm{In}_2\mathrm{O}_3$:$\mathrm{Sn}$ (ITO; $15~\Omega/\mathrm{square}$){\textemdash}for electrochemical and optical measurements{\textemdash}and onto carbon substrates for Rutherford backscattering spectrometry (RBS) analysis. Initially, the deposition system was evacuated to $\sim 10^{-7}~\mathrm{Torr}$. For removing surface contamination, the target was pre-sputtered for about $5~\mathrm{min}$ in $99.9997\%$-pure $\mathrm{Ar}$. The deposition (using a constant discharge power of $240~\mathrm{W}$) was performed in an atmosphere (at a pressure of $30~\mathrm{mTorr}$) consisting of a mixture of $\mathrm{O}_2$ ($99.998\%$-pure) and $\mathrm{Ar}$ with respective flows of $22$ and $50~\mathrm{mL}/\mathrm{min}$. Film thickness homogeneity was achieved by rotating the substrates during deposition. X-ray diffraction (XRD) patterns were obtained by a Siemens D5000 diffractometer using $\mathrm{Cu}~K\alpha$ radiation and they confirmed the amorphous nature of the $\textit{a}\mathrm{WO}_3$ films. Film thicknesses were obtained from stylus profilometry by means of a Bruker DetakXT instrument. RBS measurements were done at the Tandem Laboratory at Uppsala University employing $2~\mathrm{MeV}~^4\mathrm{He}$ ions detected at a backscattering angle of $170^\circ$. Using the SIMNRA program,\cite{Mayer1999} the elemental contents and their respective areal densities were determined by fitting the RBS spectra to a model describing the film-substrate system. The film density $\rho$ was calculated using the formula

\begin{equation}
\rho=(M_\mathrm{W}N_\mathrm{W}+M_\mathrm{O}N_\mathrm{O})/N_\mathrm{A},\label{Eq_1}
\end{equation}

with $N_\mathrm{A}$ the Avogadro constant, $M_\mathrm{W}$ ($M_\mathrm{O}$) the molar mass of tungsten (oxygen), and  $N_\mathrm{W}$ ($N_\mathrm{O}$) the number density of tungsten (oxygen). $N_\mathrm{W}$ and $N_\mathrm{O}$ correspond to the respective areal densities (obtained from RBS analysis) divided by the film thickness. 
In this work, we present representative results from a series of SECIS measurements on $\textit{a}\mathrm{WO}_3$ films produced at the same conditions, which gave qualitatively similar outcomes. The data shown here were obtained during the SECIS-1 experiment in Ref. [\onlinecite{RojasGonzalez2020}]. The studied sample was an $\textit{a}\mathrm{WO}_3$ film with thickness $d$ of $294\pm4~\mathrm{nm}$, $\mathrm{O}/\mathrm{W}$ ratio of $3.11\pm0.03$, tungsten number density $N_\mathrm{W}$ of $(1.242\pm0.002)\times10^{22}~\mathrm{cm}^{-3}$, and density $\rho$ of $4.81\pm0.02~\mathrm{g}\,\mathrm{cm}^{-3}$.

\subsection{\label{sec2:level2_2}Setup for combined electrochemical and optical measurements}

The SECIS measurement setup used in this work is described in detail elsewhere.\cite{Rojas-Gonzalez2019} Electrochemical measurements were performed in a glove box with an argon atmosphere ($\mathrm{H}_2\mathrm{O}$ level $<0.6~\mathrm{ppm}$) using a lithium-containing electrolyte{\textemdash}consisting of $1~\mathrm{M}~\mathrm{LiClO}_4$ dissolved in propylene carbonate{\textemdash}and a three-electrode setup controlled by an electrochemical interface (SI-1286, Solartron). The $\textit{a}\mathrm{WO}_3$ film was the working electrode (WE) and lithium foils acted as the counter (CE) and reference (RE) electrodes. The optical transmittance of the WE was measured using a system consisting of a LED light source (peak optical wavelength at $810~\mathrm{nm}$; M810F2, Thorlabs) and a photodetector (PDA-100A-EC, Thorlabs). The frequency-dependent measurements were carried out by a frequency response analyzer (FRA; SI-1260, Solartron) connected to the electrochemical interface. In this case, the WE is driven to a stationary equilibrium bias potential and subsequently a sinusoidal excitation voltage is superimposed to it{\textemdash}with amplitude $V_\mathrm{A}$ and circular frequency $\omega=2\pi f$ associated with the linear frequency $f$. The latter generates oscillatory current and optical transmittance responses from the electrochromic WE. The sinusoidal signals{\textemdash}that is, the voltage, current, and transmittance{\textemdash}are fed into the FRA, which measures the relative amplitudes and phases between the excitation and the responses.

\subsection{\label{sec2:level2_3} SECIS experiment}

The sequence followed for the SECIS experiment in this work is identical to that of the SECIS-1 measurement described in Ref. [\onlinecite{RojasGonzalez2020}]{\textemdash}a summary will be presented as follows and more details can be found in the given reference. The active area of the WE was $A=1.28\pm0.09~\mathrm{cm}^2$. The initial open-circuit potential of the WE (when it was just immersed in the electrolyte) was measured, and its value was $3.82~\mathrm{V}~\mathrm{vs.}~\mathrm{Li}/\mathrm{Li}^+$. Then, after cyclic voltammetry at a rate of 10 mV/s during three cycles in the potential range of $2.0$-$4.0~\mathrm{V}~\mathrm{vs.}~\mathrm{Li}/\mathrm{Li}^+$, the WE was driven to the required equilibrium bias potential by a linear potential sweep at a rate of $10~\mathrm{mV}/\mathrm{s}$. Next, the WE was maintained at the given equilibrium potential for $20~\mathrm{min}$, and afterward the SECIS measurements were made by superimposing to the equilibrium bias potential a sinusoidal voltage with amplitude $V_\mathrm{A}=20\sqrt{2}~\mathrm{mV}$. The measurements were carried out in a frequency range that spanned from $10~\mathrm{mHz}$ to $30~\mathrm{kHz}$ for equilibrium bias potential values of $3.15$, $2.90$, $2.60$, $2.10$, and $1.50~\mathrm{V}~\mathrm{vs.}~\mathrm{Li}/\mathrm{Li}^+${\textemdash}which according to figure 3(a) in Ref. [\onlinecite{RojasGonzalez2020}] corresponded respectively to the intercalation ratios $x$ (defined as $x=\mathrm{Li}/\mathrm{W}$) of $0.010$, $0.052$, $0.163$, $0.551$, and $2.013$. The total duration of the experiment was about $7.5~\mathrm{h}$.      

\subsection{\label{sec2:level2_4} Complex transfer functions}

From the experimentally obtained quantities it is possible to obtain the complex current $\tilde{I}(\omega)=I_\mathrm{A}(\omega)\, \mathrm{exp}[i\phi_I(\omega)]$ and transmittance $\tilde{T}(\omega)=T_\mathrm{A}(\omega)\, \mathrm{exp}[i\phi_\mathrm{op}(\omega)]${\textemdash}with $I_\mathrm{A}(\omega)$ and $T_\mathrm{A}(\omega)$ the respective amplitudes of the sinusoidal responses, as well as $\phi_I(\omega)$ and $\phi_\mathrm{op}(\omega)$ the relative phases with respect to that of the sinusoidal excitation voltage. The real, and imaginary part of each complex quantity corresponds respectively to the in-phase, and quadrature component, taking the excitation voltage as the reference. The electrochemical impedance $\tilde{Z}(\omega)$ is given by the ratio between the amplitude of the sinusoidal excitation voltage and the complex current, and it can be expressed as follows

\begin{equation}
\tilde{Z}(\omega)=V_\mathrm{A}/\tilde{I}(\omega)=|\tilde{Z}(\omega)|\, \mathrm{exp}[i\phi_Z(\omega)],\label{Eq_2}
\end{equation} 

with $|\tilde{Z}(\omega)|=V_\mathrm{A}/I_\mathrm{A}$ and $\phi_Z(\omega)=-\phi_I(\omega)$. We can introduce the complex charge $\tilde{Q}(\omega)$, given by

\begin{equation}
\tilde{Q}(\omega)=(i\omega)^{-1}\tilde{I}(\omega)=Q_\mathrm{A}(\omega)\, \mathrm{exp}[i\phi_C(\omega)],\label{Eq_3}
\end{equation}

with $Q_\mathrm{A}(\omega)=I_\mathrm{A}(\omega)/\omega$, and $\phi_C(\omega)=\phi_I(\omega)-\pi/2$ the amplitude, and phase of the time integral of the sinusoidal current response, respectively. By analogy with the differential capacitance, defined as the derivative of the charge with respect to the potential, the complex capacitance per unit area $\tilde{C}(\omega)$ can be defined as

\begin{equation}
\tilde{C}(\omega)=A^{-1}\tilde{Q}(\omega)/V_\mathrm{A}=|\tilde{C}(\omega)|\, \mathrm{exp}[i\phi_C(\omega)],\label{Eq_4}
\end{equation} 

with  $|\tilde{C}(\omega)|=A^{-1}Q_\mathrm{A}(\omega)/V_\mathrm{A}$ and $A$ the active area of the electrode. Note that the impedance and the complex capacitance are linked by the relation $\tilde{Z}(\omega)=A^{-1}[i\omega\tilde{C}(\omega)]^{-1}$.

The complex optical capacitance can be given by\cite{Rojas-Gonzalez2019}

\begin{equation}
\tilde{G}_\mathrm{op}(\omega)=\langle T \rangle^{-1}\tilde{T}(\omega)/V_\mathrm{A}=|\tilde{G}_\mathrm{op}(\omega)|\, \mathrm{exp}[i\phi_\mathrm{op}(\omega)],\label{Eq_5}
\end{equation}

with $|\tilde{G}_\mathrm{op}(\omega)|=\langle T \rangle^{-1}T_\mathrm{A}(\omega)/V_\mathrm{A}$ and $\langle T \rangle$ the transmittance at the corresponding equilibrium bias potential. Note that, when the variation in transmittance is much larger than that in reflectance, combining equations 3 and 5 from Ref. [\onlinecite{RojasGonzalez2020}] it is possible to show that a small variation in the optical density (defined as the optical absorption coefficient multiplied by the film thickness) is proportional to a term analogous to $\langle T \rangle^{-1}\tilde{T}(\omega)$. So, the quantity $\langle T \rangle^{-1}\tilde{T}(\omega)${\textemdash}related to the oscillatory changes of the optical absorption{\textemdash}could be thought as an "optical charge," and this emphasizes the analogy between  $\tilde{C}(\omega)$ and $\tilde{G}_\mathrm{op}(\omega)$.  

\subsection{\label{sec2:level2_5} Equivalent circuit for fitting the EIS spectra}

\begin{figure*}[ht!]
\includegraphics[scale=0.94]{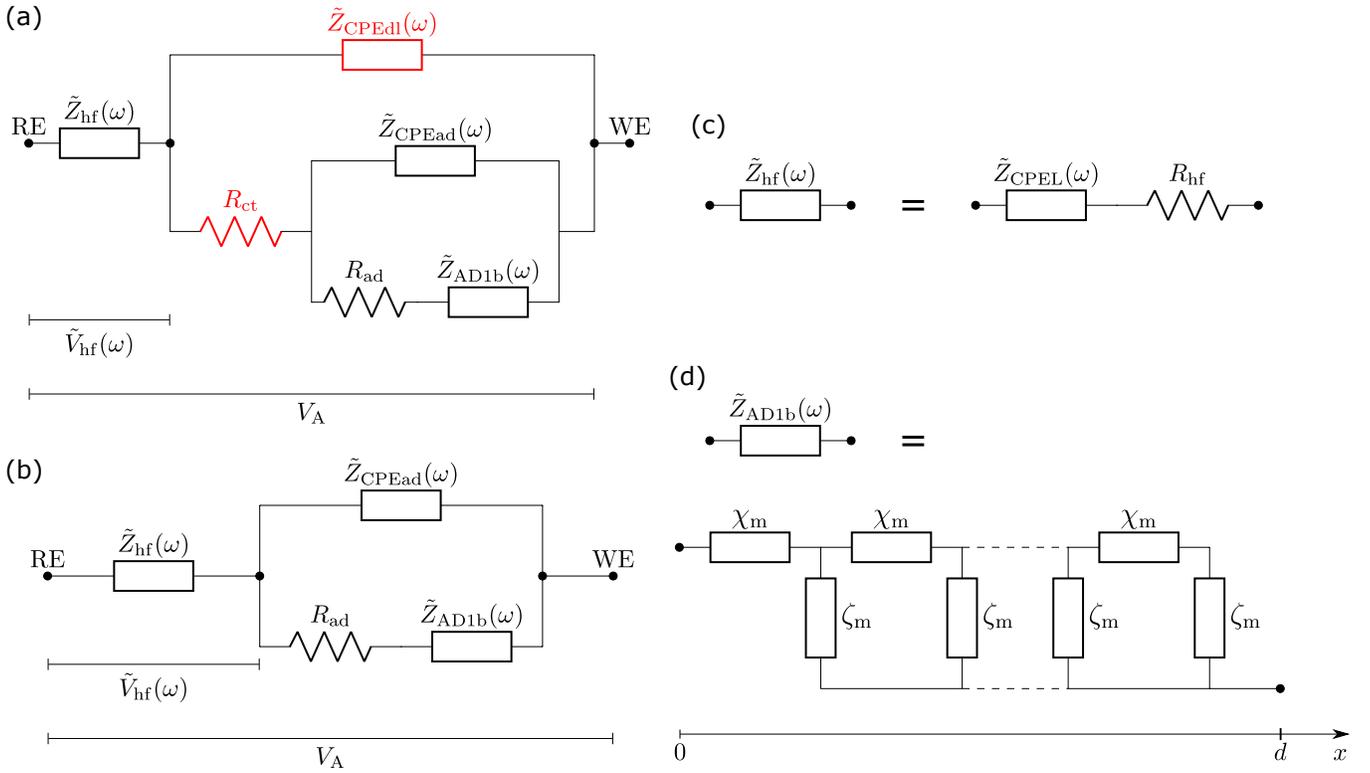}
\caption{\label{fig:1} (a) Complete equivalent circuit used for the fitting of the EIS spectra. Here, the total amplitude of the sinusoidal excitation voltage $V_\mathrm{A}$ is applied between the WE and the RE. This circuit consists of a high-frequency impedance $\tilde{Z}_\mathrm{hf}(\omega)$, with an associated voltage drop $\tilde{V}_\mathrm{hf}(\omega)$, in series with a combination that describes the impedance at the interface between the electrolyte and the EC film (the elements of this combination are described in the main text). (b) Simplified form of the equivalent circuit presented in (a), without the elements $R_\mathrm{ct}$ and $\tilde{Z}_\mathrm{CPEdl}(\omega)$, valid when the double layer gives a negligible contribution to the total impedance. (c) Expansion of $\tilde{Z}_\mathrm{hf}(\omega)$ showing that it is composed of an inductive constant phase element $\tilde{Z}_\mathrm{CPEL}(\omega)$ in series with a high-frequency resistance $R_\mathrm{hf}$. (d) Transmission line of length $d$ with the same impedance as that of the anomalous diffusion element $\tilde{Z}_\mathrm{AD1b}(\omega)$ from Eq. (\ref{Eq_6}){\textemdash}the elements with impedances $\chi_\mathrm{m}$ and $\zeta_\mathrm{m}$ are described in the main text.}
\end{figure*}

The equivalent circuit used for fitting the EIS spectra is depicted in Fig. \ref{fig:1}. The complete circuit is shown in Fig. \ref{fig:1}(a), and it consists of a series array of the uncompensated high-frequency impedance $\tilde{Z}_\mathrm{hf}(\omega)$, that at the electrolyte-$\textit{a}\mathrm{WO}_3$ film interface, and the diffusion impedance in the film. The term $\tilde{Z}_\mathrm{hf}(\omega)$, expanded in Fig. \ref{fig:1}(c), contains a high-frequency resistance $R_\mathrm{hf}$ (accounting for the electrolyte, the ITO, and the connections to the electrochemical cell) in series with an inductive constant phase element (CPE) $\tilde{Z}_\mathrm{CPEL}(\omega)=[Q_\mathrm{L}(i\omega)^{n_\mathrm{L}}]^{-1}$ mainly due to the connection of the cables to the cell. The interfacial impedance shown in Fig. 1(a) comprises a double layer CPE $\tilde{Z}_\mathrm{CPEdl}(\omega)=[Q_\mathrm{dl}(i\omega)^{n_\mathrm{dl}}]^{-1}$ in parallel with a branch containing a charge transfer resistance $R_\mathrm{ct}$ in series with a parallel array describing an adsorption step\cite{Franceschetti1982} and the chemical diffusion of the ions and the electrons in the bulk of the EC film. The latter consists of an adsorption CPE $\tilde{Z}_\mathrm{CPEad}(\omega)=[Q_\mathrm{ad}(i\omega)^{n_\mathrm{ad}}]^{-1}$ in parallel with a branch given by an adsorption resistance $R_\mathrm{ad}$ in series with a diffusion element $\tilde{Z}_\mathrm{AD1b}(\omega)$. According to Ref. [\onlinecite{Franceschetti1979}], the adsorption elements{\textemdash}$R_\mathrm{ad}$ and $\tilde{Z}_\mathrm{CPEad}(\omega)${\textemdash}are related to each other and depend on $R_\mathrm{ct}$. The adsorption parameters describe the effect on the intercalation rate of the formation of ion-electron pairs at the electrolyte-EC film interface. The ion diffusion is described by the anomalous diffusion 1b (AD1b) model with reflecting boundary condition presented in Ref. [\onlinecite{Bisquert2001}] and with impedance given by

\begin{equation}
\tilde{Z}_\mathrm{AD1b}(\omega)=R_\mathrm{W}\omega_\mathrm{D}^{\gamma-1}[\omega_\mathrm{D}/(i\omega)]^{1-\gamma/2}\,\mathrm{coth}[(i\omega/\omega_\mathrm{D})^{\gamma/2}],\label{Eq_6}
\end{equation} 

with $R_\mathrm{W}$ a multiplicative factor, $\gamma$ the anomalous diffusion exponent, and the characteristic frequency of diffusion given by

\begin{equation}
\omega_\mathrm{D}=2\pi f_\mathrm{D}=(AR_\mathrm{W}C_\mathrm{ch})^{-1/\gamma},\label{Eq_7}
\end{equation} 

with $f_\mathrm{D}$ the characteristic linear frequency of diffusion and $C_\mathrm{ch}$ the diffusion capacitance (per unit area), which can also be considered as a chemical capacitance.\cite{Bisquert2003a} An effective chemical diffusion coefficient $D_\mathrm{ch}$ (in ordinary units of $\mathrm{cm}^2\,\mathrm{s}^{-1}$) can be expressed as\cite{Backholm2008,Malmgren2017}

\begin{equation}
D_\mathrm{ch}=d^2 \omega_\mathrm{D},\label{Eq_8}
\end{equation} 

Note that the diffusion coefficient given in Ref. [\onlinecite{Bisquert2001}] (within the AD1b model) does not have the typical dimensions ($\mathrm{cm}^2\,\mathrm{s}^{-1}$) associated with diffusion.

Regarding the anomalous diffusion element, the impedance expression in Eq. (\ref{Eq_6}) is equivalent (mathematically) to that of the transmission line with open-circuit termination depicted in Fig. \ref{fig:1}(d). The latter has a length $d$ and is composed of an array of elements with scaled impedances $\chi_\mathrm{m}$ [$\Omega\,\mathrm{m}^{-1}$] and $\zeta_\mathrm{m}$ [$\Omega\,\mathrm{m}$]. For the AD1b model, $\chi_\mathrm{m}$ corresponds to a CPE and $\zeta_\mathrm{m}$ to a capacitor with

\begin{equation}
\chi_\mathrm{m}=[Q_\mathrm{m}(i\omega)^{n_\mathrm{m}}]^{-1},\label{Eq_9}
\end{equation} 

\begin{equation}
\zeta_\mathrm{m}=(i\omega c_\mathrm{m})^{-1}.\label{Eq_10}
\end{equation} 

The parameters used in the fitting procedure were $Q_\mathrm{m}$, $n_\mathrm{m}$, and $c_\mathrm{m}$. Their connection with Eqs. (\ref{Eq_6}) and (\ref{Eq_7}) is determined by the relations $R_\mathrm{W}=d/Q_\mathrm{m}$, $C_\mathrm{ch}=A^{-1} c_\mathrm{m} d$, and $\gamma=1-n_\mathrm{m}$. The physics behind the derivation of the $\tilde{Z}_\mathrm{AD1b}(\omega)$ element is related to the continuous-time random walk model.\cite{Bisquert2001,Bisquert2003} It has been used before for fitting the EIS spectra of $\textit{a}\mathrm{WO}_3$ films, showing satisfactory results within the bias potential range used in this work.\cite{Malmgren2017} Alternatively, the impedance related to the anomalous diffusion 1a (AD1a) model with reflecting boundary condition from Ref. [\onlinecite{Bisquert2001}] could be an option. The AD1a model corresponds to a case in which the number of ions taking part of the diffusion is not conserved{\textemdash}this occurs, for example, during irreversible ion trapping. When applied to $\textit{a}\mathrm{WO}_3$ in a previous work, it provided better fitting results than the AD1b model for potentials lower than about $1.8~\mathrm{V}~\mathrm{vs}.~\mathrm{Li}/\mathrm{Li}^+$.\cite{Malmgren2017} In the present work, both the AD1a and the AD1b models were tested. The latter gave the best results in the fittings for all the studied bias potentials. It is worth mentioning that the normal diffusion is a special case of the anomalous diffusion models{\textemdash}namely, it is obtained for $\gamma=1$.

Note that, in the equivalent circuit outlined in Fig. \ref{fig:1}, the elements corresponding to the capacitances and inductances are presented in the general constant phase element form. In this case, perfect capacitances are retrieved when the exponents $n_\mathrm{dl}$ and $n_\mathrm{ad}$ are equal to $1$. Similarly, $n_\mathrm{L}=-1$ corresponds to a perfect inductance. The effective double layer and adsorption capacitances (per unit area), denoted respectively by $C_\mathrm{dl}$ and $C_\mathrm{ad}$, can be estimated{\textemdash}assuming the normal time-constant distribution case presented in Ref. [\onlinecite{Hirschorn2010}]{\textemdash}using the formulas given by

\begin{equation}
C_\mathrm{dl}=A^{-1}Q_\mathrm{dl}^{1/n_\mathrm{dl}}R_\mathrm{ct}^{(1-n_\mathrm{dl})/n_\mathrm{dl}},\label{Eq_11}
\end{equation}

\begin{equation}
C_\mathrm{ad}=A^{-1}Q_\mathrm{ad}^{1/n_\mathrm{ad}}R_\mathrm{ad}^{(1-n_\mathrm{ad})/n_\mathrm{ad}}.\label{Eq_12}
\end{equation}

The effective characteristic frequency of the double layer response $\omega_\mathrm{dl}=2\pi f_\mathrm{dl}$ can be expressed as

\begin{equation}
\omega_\mathrm{dl}=(AR_\mathrm{ct}C_\mathrm{dl})^{-1}.\label{Eq_13}
\end{equation}

Similarly, the effective characteristic frequency of the adsorption process $\omega_\mathrm{ad}=2\pi f_\mathrm{ad}$ takes the form

\begin{equation}
\omega_\mathrm{ad}=(AR_\mathrm{ad}C_\mathrm{ad})^{-1}.\label{Eq_14}
\end{equation}

Figure \ref{fig:1}(b) portrays a simplified case for $f\ll f_\mathrm{dl}$. Here, we can neglect the contribution to the total impedance made by the element $\tilde{Z}_\mathrm{CPEdl}(\omega)$, and the effect of $R_\mathrm{ct}$ (presumably small) would be absorbed by the high-frequency resistance $R_\mathrm{hf}$. To give an idea of the associated orders of magnitude, assuming a typical double layer capacitance value for a semiconductor WE of
$C_\mathrm{dl}=10^{-5}~\mathrm{F}\, \mathrm{cm}^{-2}$,\cite{Hankin2019} using $R_\mathrm{ct}=1~\Omega$ we get $f_\mathrm{dl}\approx 13~\mathrm{kHz}$. Likewise, for $R_\mathrm{ct}=5~\Omega$, we end up with $f_\mathrm{dl}\approx 2.5~\mathrm{kHz}$.

Focusing on the parallel array describing the adsorption and diffusion (outlined in Fig. \ref{fig:1}), the adsorption branch consists of the element $\tilde{Z}_\mathrm{CPEad}(\omega)$ and its related complex capacitance $\tilde{C}_\mathrm{CPEad}(\omega)$ is given by

\begin{equation}
\tilde{C}_\mathrm{CPEad}(\omega)=A^{-1}[i\omega\tilde{Z}_\mathrm{CPEad}(\omega)]^{-1}.\label{Eq_15}
\end{equation}

Similarly, the branch describing the intercalation is composed of the series combination $R_\mathrm{ad}$-$\tilde{Z}_\mathrm{AD1b}(\omega)$, and its associated complex capacitance $\tilde{C}_\mathrm{int}(\omega)$ takes the form

\begin{equation}
\tilde{C}_\mathrm{int}(\omega)=A^{-1}\{i\omega[R_\mathrm{ad}+\tilde{Z}_\mathrm{AD1b}(\omega)]\}^{-1}.\label{Eq_16}
\end{equation}

The complex Faradaic capacitance $\tilde{C}_\mathrm{F}(\omega)$, accounting only for the voltage drop at the electrolyte-EC film interface, can be defined as

\begin{equation}
\tilde{C}_\mathrm{F}(\omega)=\tilde{C}_\mathrm{CPEad}(\omega)+\tilde{C}_\mathrm{int}(\omega).\label{Eq_17}
\end{equation}

Here, it was assumed that the double layer capacitance was much smaller than those of the adsorption and intercalation branches, and that both $\tilde{C}_\mathrm{CPEad}(\omega)$ and $\tilde{C}_\mathrm{int}(\omega)$ could contribute to the Faradaic capacitance. 

The quantity $\tilde{G}_\mathrm{op}(\omega)$, defined in Eq. (\ref{Eq_5}), describes the optical behavior of the whole EC system because it considers the total voltage drop between the WE and the RE. A modified complex optical capacitance $\tilde{G}_\mathrm{op}^\mathrm{F}(\omega)${\textemdash}considering only the voltage drop related to the Faradaic processes{\textemdash}can be expressed as

\begin{eqnarray}
\tilde{G}_\mathrm{op}^\mathrm{F}(\omega)&=&\langle T \rangle^{-1}\tilde{T}(\omega)/[V_\mathrm{A}-\tilde{V}_\mathrm{hf}(\omega)] \nonumber\\ 
&=&\tilde{G}_\mathrm{op}(\omega)[1-\tilde{Z}_\mathrm{hf}(\omega)/\tilde{Z}_\mathrm{fit}(\omega)]^{-1},\label{Eq_18}
\end{eqnarray}

with $\tilde{Z}_\mathrm{fit}(\omega)$ the total impedance obtained from the fittings. The core of the present work is based on the comparison between $\tilde{G}_\mathrm{op}^\mathrm{F}(\omega)$ and the quantities $\tilde{C}_\mathrm{F}(\omega)$, $\tilde{C}_\mathrm{CPEad}(\omega)$, and $\tilde{C}_\mathrm{int}(\omega)$.

\section{\label{se3:level1}Results}

\begin{figure*}[ht!]
\includegraphics[scale=0.36]{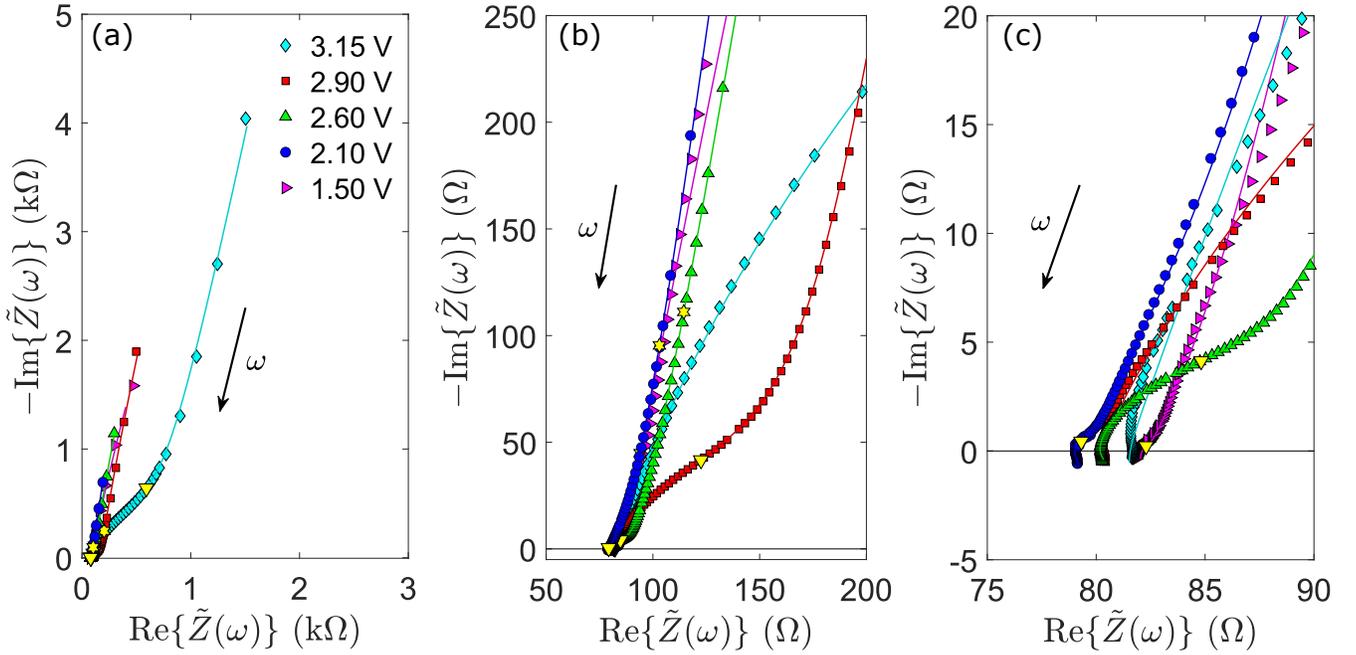}
\caption{\label{fig:2} (a) Nyquist plot of the EIS spectra for the $\textit{a}\mathrm{WO}_3$ film at the equilibrium bias potentials (with respect to $\mathrm{Li}/\mathrm{Li}^+$) given in the plot. The symbols depict the experimental data, and the lines correspond to the fittings to the equivalent circuit portrayed in Fig. \ref{fig:1}(b). Expanded views of (a) are presented in (b) and (c). The arrows indicate the direction of increasing frequency $\omega$. The points corresponding to the linear frequencies $f_\mathrm{ad}$ (yellow downward-facing triangles) and $f_\mathrm{D}$ (yellow stars) are marked. The frequency $f_\mathrm{D}$ at $3.15~\mathrm{V}~\mathrm{vs}.~\mathrm{Li}/\mathrm{Li}^+$ is outside the experimental frequency range. Here, the spectra are displayed in the frequency range between $10~\mathrm{mHz}$ and $10~\mathrm{kHz}$. Figure S1 in the Supplementary Information (SI) depicts each EIS spectrum in the frequency range of $10~\mathrm{mHz}$-$30~\mathrm{kHz}$.}
\end{figure*}

\begin{table*}[ht!]
\caption{\label{table:I} Parameters obtained from the fitting of the experimental EIS spectra shown in Fig. \ref{fig:2} to the equivalent circuit depicted in Fig. \ref{fig:1}(b). For all the parameters which were let free during the fittings, the fitting error is indicated between round brackets. In addition, the $\chi^2$ value of each fit is presented. }
\begin{ruledtabular}
\begin{tabular}{>{\centering\arraybackslash}m{1.8cm}>{\centering\arraybackslash}m{1.0cm}>{\centering\arraybackslash}m{1.1cm}>{\centering\arraybackslash}m{1.0cm}>{\centering\arraybackslash}m{1.0cm}>{\centering\arraybackslash}m{1.1cm}>{\centering\arraybackslash}m{1.0cm}>{\centering\arraybackslash}m{1.7cm}>{\centering\arraybackslash}m{1.0cm}>{\centering\arraybackslash}m{1.6cm}>{\centering\arraybackslash}m{1.6cm}}
Bias potential & $R_\mathrm{hf}$ & $Q_\mathrm{L}$ & $n_\mathrm{L}$ & $Q_\mathrm{ad}$ & $n_\mathrm{ad}$ & $R_\mathrm{ad}$ & $Q_\mathrm{m}$ & $n_\mathrm{m}$ & $c_\mathrm{m}$ & $\chi^2$\\
($\mathrm{V}~\mathrm{vs}.~\mathrm{Li}/\mathrm{Li}^+$)& ($\Omega$)& ($\mathrm{F}\,\mathrm{s}^{n_\mathrm{L}-1}$) & & ($\mu\mathrm{F}\,\mathrm{s}^{n_\mathrm{ad}-1}$) &  & ($\Omega$) & ($\mathrm{F}\,\mathrm{s}^{n_\mathrm{m}-1}\,\mathrm{m}$) & & ($\mathrm{F}$ $\mathrm{m}^{-1}$) & \\ \hline
$3.15$ & $81.45$ & $89462$ & $-1$ & $890$ & $0.785$ & $1353$ & $2.2\times 10^{-10}$ & $0.62$ & $11006$ & $2.57\times 10^{-4}$\\
 & ($0.07\%$) & ($25.75\%$) & & ($0.55\%$) & ($0.20\%$) & ($2.03\%$) & ($4.89\%$) & ($5.54\%$) & ($5.90\%$) & \\
& & & & & & & & & & \\ 
$2.90$ & $80.1$ & $71013$ & $-1$ & $1592$ & $0.725$ & $96.08$ & $1.077\times 10^{-9}$ & $0.279$ & $17493$ & $8.46\times 10^{-5}$\\
 & ($0.04\%$) & ($10.33\%$) & & ($0.63\%$) & ($0.18\%$) & ($0.99\%$) & ($0.83\%$) & ($2.46\%$) & ($0.69\%$) & \\ 
& & & & & & & & & & \\ 
$2.60$ & $79.64$ & $182.1$ & $-0.45$ & $2185$ & $0.717$ & $11.18$ & $2.66\times 10^{-9}$ & $0.448$ & $33758$ & $6.27\times 10^{-5}$\\
 & ($0.19\%$) & ($66\%$) & ($18.86\%$) & ($1.57\%$) & ($0.42\%$) & ($1.10\%$) & ($1.00\%$) & ($1.18\%$) & ($0.65\%$) & \\ 
& & & & & & & & & & \\ 
$2.10$ & $78.75$ & $1447$ & $-0.64$ & $3308$ & $0.75$ & $1.20$ & $3.67\times 10^{-9}$ & $0.404$ & $60352$ & $1.57\times 10^{-4}$\\
 & ($0.10\%$) & ($10.18\%$) & & ($9.70\%$) & ($2.02\%$) & ($6.32\%$) & ($2.31\%$) & ($1.37\%$) & ($2.72\%$) & \\ 
& & & & & & & & & & \\ 
$1.50$ & $82.0$ & $130680$ & $-1$ & $2410$ & $0.712$ & $0.75$ & $3.46\times 10^{-9}$ & $0.458$ & $21321$ & $8.24\times 10^{-3}$\\
 & ($0.31\%$) & ($1.37\%$) & & ($2.55\%$) & ($0.53\%$) & ($1.64\%$) & ($4.60\%$) & ($1.90\%$) & ($2.07\%$) & \\ 
\end{tabular}
\end{ruledtabular}
\end{table*}

\begin{figure*}[ht!]
\includegraphics[scale=0.27]{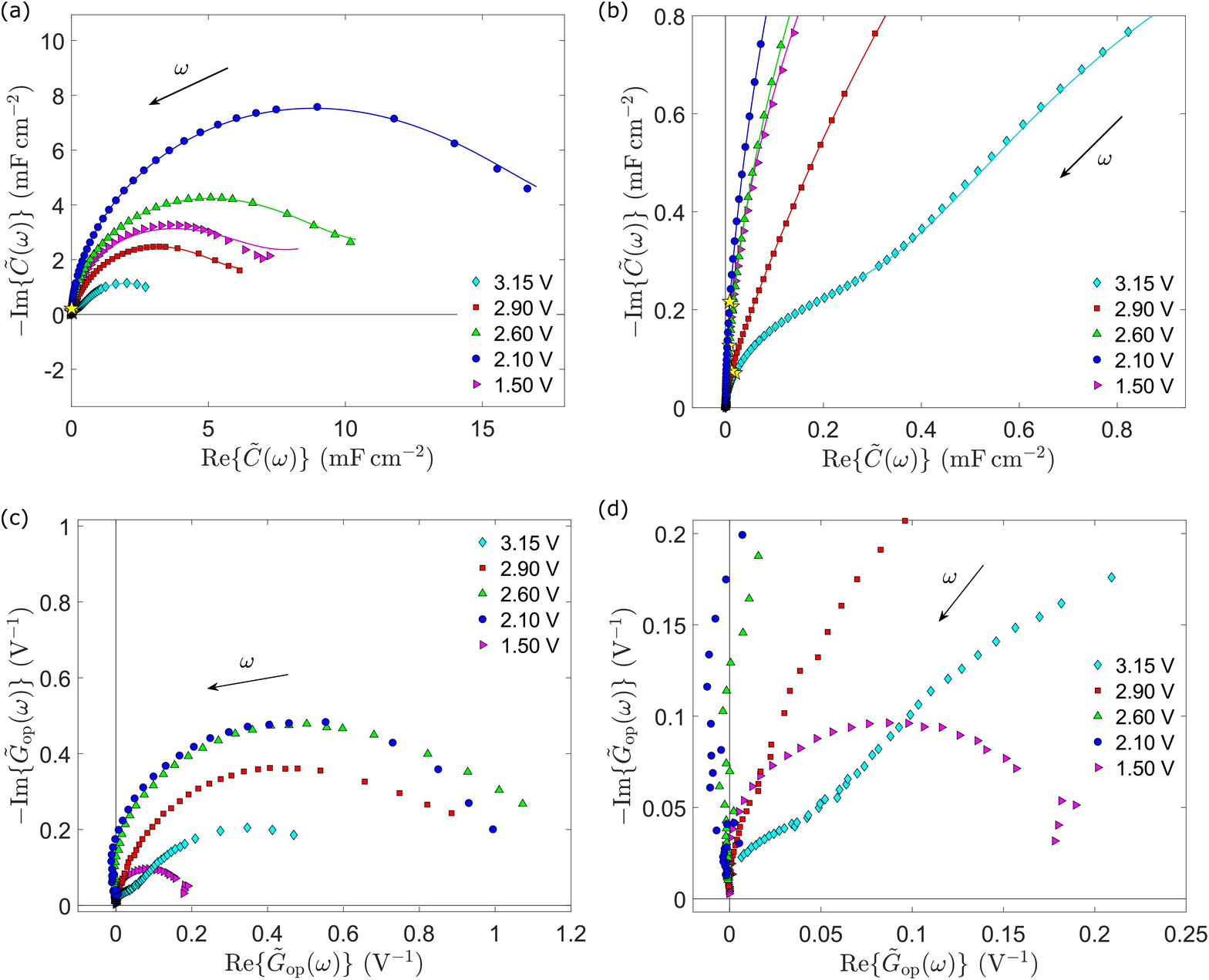}
\caption{\label{fig:3} Nyquist plots of the total complex capacitance $\tilde{C}(\omega)$ (a) and the complex optical capacitance $\tilde{G}_\mathrm{op}(\omega)$ (c) for the $\textit{a}\mathrm{WO}_3$ film measured at the bias potentials, with respect to $\mathrm{Li}/\mathrm{Li}^+$, given in the plots. Note that both $\tilde{C}(\omega)$ and $\tilde{G}_\mathrm{op}(\omega)$ contain the effect of the high-frequency impedance $\tilde{Z}_\mathrm{hf}(\omega)$. Panels (b), and (d) display the high-frequency region of panels (a), and (c), respectively. The frequency $\omega$ increases at the direction specified by the arrows. The symbols correspond to the experimental data and the lines, in (a) as well as in (b), represent the fittings to the equivalent circuit depicted in Fig. \ref{fig:1}(b). In (a), the spectra are presented for the frequency range of $10~\mathrm{mHz}$-$10~\mathrm{kHz}$. In (c), the lowest displayed frequency was $10~\mathrm{mHz}$ and the highest frequencies were about $18$, $18$, $11$, $7$, and $7~\mathrm{Hz}$ for $3.15$, $2.90$, $2.60$, $2.10$, and $1.50~\mathrm{V}~\mathrm{vs}.~\mathrm{Li}/\mathrm{Li}^+$, respectively. The upper limits were chosen in such a way that the data with signal-to-noise ratio of the optical signal lower than $5$ were not shown{\textemdash}the data points corresponding to these frequencies are indicated in (a) and (b) by yellow stars. Figure S3 in the SI portrays the Bode representation of $\tilde{C}(\omega)$ (in this case, including frequencies up to $30~\mathrm{kHz}$) and $\tilde{G}_\mathrm{op}(\omega)$.}
\end{figure*}

\begin{table*}[t]
\caption{\label{table:II} Characteristic frequencies and effective capacitances of the adsorption and diffusion processes as well as the effective diffusion coefficients for the $\textit{a}\mathrm{WO}_3$ film at the studied equilibrium bias potentials (with the associated intercalation levels $x$). The given quantities were calculated using the fitting parameters presented in Table \ref{table:I}. For the sake of comparison, for $C_\mathrm{ad}$ and $C_\mathrm{ch}$, the errors excluding the contribution of that of the active area $A$ are presented in between round brackets. Note that $A$ only gives a multiplicative factor that is common for all bias potentials{\textemdash}in this case, it usually represents the largest (or one of the largest) sources of error. }
\begin{ruledtabular}
\begin{tabular}{>{\centering\arraybackslash}m{1.8cm}>{\centering\arraybackslash}m{1.4cm}>{\centering\arraybackslash}m{1.8cm}>{\centering\arraybackslash}m{1.8cm}>{\centering\arraybackslash}m{1.8cm}>{\centering\arraybackslash}m{2.4cm}>{\centering\arraybackslash}m{2.6cm}}
Bias potential & $x=\mathrm{Li}/\mathrm{W}$ & $f_\mathrm{ad}$ & $C_\mathrm{ad}$ & $f_\mathrm{D}$ & $C_\mathrm{ch}$ & $D_\mathrm{ch}$ \\
$(\mathrm{V}~\mathrm{vs}.~\mathrm{Li}/\mathrm{Li}^+)$ &  & ($\mathrm{Hz}$) & ($\mu\mathrm{F}\, \mathrm{cm}^{-2}$) & ($\mathrm{Hz}$) & ($\mathrm{mF}\, \mathrm{cm}^{-2}$) & ($\mathrm{s}^{-1}\,\mathrm{cm}^2$) \\ \hline
3.15 & $0.010$ & $0.126\pm 0.003$ & $731\pm 52(7)$ & $0.003\pm 0.001$ & $2.53\pm 0.23(0.15)$ & $(1.7\pm 0.7)\times 10^{-11}$ \\
2.90 & $0.052$ & $2.12\pm 0.04$ & $609\pm 43(6)$ & $0.100\pm 0.004$ & $4.01\pm 0.29(0.06)$ & $(5.4\pm 0.1)\times 10^{-10}$ \\
2.60 & $0.163$ & $28\pm 1$ & $395\pm 30(12)$ & $0.135\pm 0.007$ & $7.75\pm 0.56(0.12)$ & $(7.3\pm 0.2)\times 10^{-10}$ \\
2.10 & $0.551$ & $246\pm 53$ & $423\pm 89(84)$ & $0.089\pm 0.007$ & $13.85\pm 1.06(0.42)$ & $(4.8\pm 0.3)\times 10^{-10}$ \\
1.50 & $2.013$ & $1133\pm 72$ & $146\pm 13(9)$ & $0.51\pm 0.05$ & $4.89\pm 0.36(0.12)$ & $(2.8\pm 0.3)\times 10^{-9}$ \\
\end{tabular}
\end{ruledtabular}
\end{table*}

Figure \ref{fig:2} depicts the complex plane representation of the experimental impedance spectra for the $\textit{a}\mathrm{WO}_3$ film. Detailed plots for each bias potential are shown in Fig. S1 in the Supplementary Information (SI). The spectra show the same qualitative shape. That is, from high to low frequencies, a depressed semicircle followed by a diffusion region which (when shown within the displayed frequency range) makes an upturn toward a blocking electrode response that, in this case, is inclined with respect to the vertical axis.

In general, the lower the bias potential the lower the impedance values, only the $1.50~\mathrm{V}~\mathrm{vs}.~\mathrm{Li}/\mathrm{Li}^+$ spectrum shows low-frequency values higher than those from the $2.10$ and $2.60~\mathrm{V}~\mathrm{vs}.~\mathrm{Li}/\mathrm{Li}^+$ data sets. In addition, high-frequency data points with $\mathrm{Im}\{\tilde{Z}(\omega)\}>0$, representing inductive effects mostly due to the connection cables, can be observed in Fig. \ref{fig:2}(c) and Fig. S1 in the SI{\textemdash}the latter showing a higher detail. These inductive effects are small{\textemdash}that is, the imaginary parts of the spectra present a value of about $2~\Omega$ at $30~\mathrm{kHz}$. However, their relative importance is higher for the lower bias potentials because the size of the depressed semicircle features decreases with decreasing bias potential. It is worth mentioning that, in all cases, only one clear semicircle is distinguished. This justifies the use (in the fitting procedure) of the equivalent circuit depicted in Fig. \ref{fig:1}(b). Here, the semicircle is assigned to a parallel R-CPE combination related to an adsorption step at the interface between the electrolyte and the film. Moreover, the absence of a visible feature linked to the double layer, in the spectra shown in Fig. \ref{fig:2}, could be due to a very small $R_\mathrm{ct}$ value, see Fig. \ref{fig:1}(a). The latter, on top of resulting in an insignificantly small semicircle in the complex plane, would push the associated characteristic frequency toward high values, see Eq. (\ref{Eq_13}), and this would hide the double layer effect below the inductive contributions.

In general, the fittings of the EIS spectra to the equivalent circuit shown in Fig. \ref{fig:1}(b) are excellent{\textemdash}see Fig. \ref{fig:2}. However, for the $1.50~\mathrm{V}~\mathrm{vs}.~\mathrm{Li}/\mathrm{Li}^+$ case, a good fitting could not be completely achieved, see the clear deviations between the line and the symbols (belonging to this bias potential) depicted in Fig. \ref{fig:2} as well as in Figs. S1(i) and S1(j) in the SI. It has been pointed out before that, at low bias potentials (below about $2~\mathrm{V}~\mathrm{vs}.~\mathrm{Li}/\mathrm{Li}^+$), degradation processes start to occur in the $\textit{a}\mathrm{WO}_3$ film and that the stationary condition requirement is not fully achieved during frequency-dependent measurements.\cite{RojasGonzalez2020} These processes could introduce effects in the EIS spectrum that cannot be fully reconciled with an equivalent circuit which models a stable system.

The fitting parameters for the different bias potentials are presented in Table \ref{table:I}. The $\chi^2$ values, describing the goodness of the fit (the lower the better), are of the same order of magnitude in the bias potential range of $2.10$-$3.15~\mathrm{V}~\mathrm{vs}.~\mathrm{Li}/\mathrm{Li}^+$, presenting the lowest value at $2.60~\mathrm{V}~\mathrm{vs}.~\mathrm{Li}/\mathrm{Li}^+$. As expected from the discussion around Fig. \ref{fig:2}, the $\chi^2$ value for the $1.50~\mathrm{V}~\mathrm{vs}.~\mathrm{Li}/\mathrm{Li}^+$ case was about two orders of magnitude higher than the rest. The high-frequency resistance $R_\mathrm{hf}$ is practically constant during the whole series of measurements, presenting a value of about $80.4\pm1.7~\Omega$. Perfect inductors (with $n_\mathrm{L}=-1$) were used except for the bias potentials of $2.10$ and $2.60~\mathrm{V}~\mathrm{vs}.~\mathrm{Li}/\mathrm{Li}^+$. In these cases, an inductive constant phase element with $n_\mathrm{L}$ different than $-1$ gave significantly better results{\textemdash}that is, the fitting to the depressed semicircle feature presented a substantial improvement and the $\chi^2$ value decreased with respect to the $n_\mathrm{L}=-1$ situation. A good agreement between the experimental data and the models is needed for the calculation of the Faradaic quantities{\textemdash}that is $\tilde{C}_\mathrm{F}(\omega)$ and $\tilde{G}_\mathrm{op}^\mathrm{F}(\omega)${\textemdash}and for being able to make meaningful comparisons between them. Thus, the $\tilde{Z}_\mathrm{CPEL}(\omega)$ element was necessary for correcting the distortions caused by the inductive effects, mainly on the depressed semicircles and their high-frequency crossing with the horizontal axis. The $R_\mathrm{ad}$ parameter is related to how difficult it is for the ions to get into the bulk of the film. Graphically, in Fig. \ref{fig:2}, it represents the size of the chord formed between the intersections of the depressed semicircle (linked here to the adsorption) and the horizontal axis. The value of $R_\mathrm{ad}$ diminishes drastically with decreasing bias potential. In fact, it changes from $1353$ to $11.18~\Omega$ between $3.15$ and $2.60~\mathrm{V}~\mathrm{vs}.~\mathrm{Li}/\mathrm{Li}^+$ and decreases further to values of the order of $1~\Omega$ at lower bias potentials. 

The remaining parameters presented in Table \ref{table:I} can be better analyzed by looking at the derived quantities shown in Table \ref{table:II}. The characteristic frequency of the adsorption process, $f_\mathrm{ad}$, increases with decreasing bias potential. The effective adsorption capacitance, $C_\mathrm{ad}$, decreases from $3.15$ down to $2.60~\mathrm{V}~\mathrm{vs}.~\mathrm{Li}/\mathrm{Li}^+$, stabilizes between $2.60$ and $2.10~\mathrm{V}~\mathrm{vs}.~\mathrm{Li}/\mathrm{Li}^+$, and decreases again at $1.50~\mathrm{V}~\mathrm{vs}.~\mathrm{Li}/\mathrm{Li}^+$. Notably, $C_\mathrm{ad}$ presents values between $100$ and $1000~\mu\mathrm{F}~\mathrm{cm}^{-2}$, and these figures are much higher than the typical double layer capacitance values ($\sim10~\mu\mathrm{F}~\mathrm{cm}^{-2}$). These values contribute to the argument of assigning the observable semicircle features (in Fig. \ref{fig:2}) to the adsorption process and not to the double layer. The chemical capacitance $C_\mathrm{ch}$ presents values between about $2$ to $15~\mathrm{mF}~\mathrm{cm}^{-2}$, increasing monotonously from $3.15$ down to $2.10~\mathrm{V}~\mathrm{vs}.~\mathrm{Li}/\mathrm{Li}^+$ and showing a subsequent decrease at $1.50~\mathrm{V}~\mathrm{vs}.~\mathrm{Li}/\mathrm{Li}^+$. The characteristic linear frequency of diffusion $f_\mathrm{D}$ and the chemical diffusion coefficient $D_\mathrm{ch}$ are linked by Eq. (\ref{Eq_8}), both show their lowest and highest value at $3.15$ and $1.50~\mathrm{V}~\mathrm{vs}.~\mathrm{Li}/\mathrm{Li}^+$, respectively. In addition, their respective values remain of the same order of magnitude between $2.90$ and $2.10~\mathrm{V}~\mathrm{vs}.~\mathrm{Li}/\mathrm{Li}^+$. $D_\mathrm{ch}$ presents values in the range of about $10^{-11}$-$10^{-9}~\mathrm{s}^{-1}\mathrm{cm}^2$. Figure S2 in the SI depicts a comparison between the bias potential dependence of $C_\mathrm{ad}$, $C_\mathrm{ch}$, and $D_\mathrm{ch}$.

\begin{figure*}[htp!]
\includegraphics[scale=0.185]{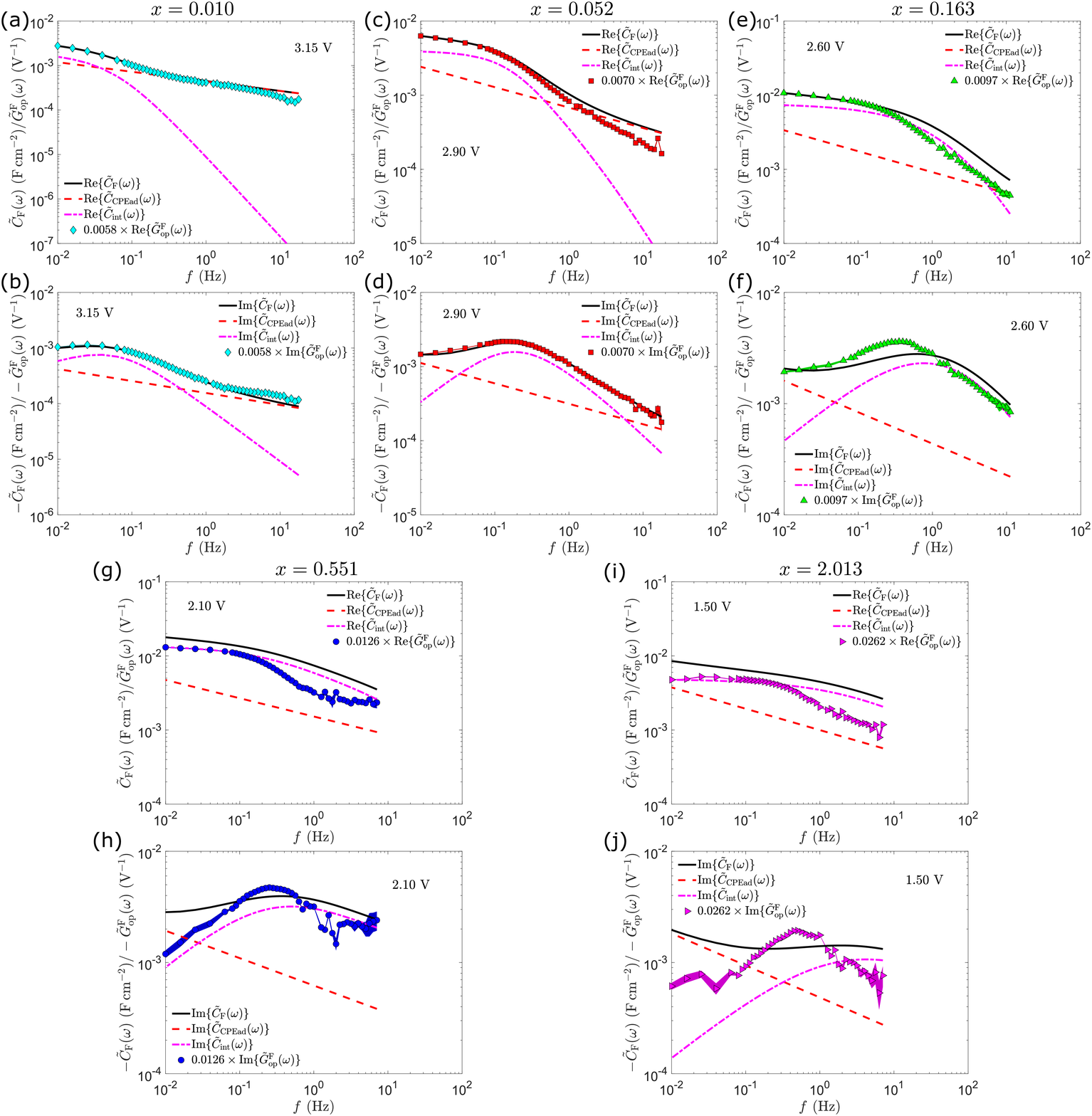}
\caption{\label{fig:4} Bode plots of the real{\textemdash}(a), (c), (e), (g), and (i){\textemdash}and imaginary{\textemdash}(b), (d), (f), (h), and (j){\textemdash}parts of the complex Faradaic capacitance $\tilde{C}_\mathrm{F}(\omega)$, solid black lines, and optical capacitance $\tilde{G}_\mathrm{op}^\mathrm{F}(\omega)$, symbols, for the $\textit{a}\mathrm{WO}_3$ film corresponding to the bias potential values (with respect to $\mathrm{Li}/\mathrm{Li}^+$) of $3.15~\mathrm{V}$ [(a) and (b)], $2.90~\mathrm{V}$ [(c) and (d)], $2.60~\mathrm{V}$ [(e) and (f)], $2.10~\mathrm{V}$ [(g) and (h)], and $1.50~\mathrm{V}$ [(i) and (j)]. The intercalation level $x$ associated with each bias potential is indicated on top of each pair of plots related to the same equilibrium condition. $\tilde{C}_\mathrm{F}(\omega)$ and $\tilde{G}_\mathrm{op}^\mathrm{F}(\omega)$ were calculated{\textemdash}employing Eqs. (\ref{Eq_17}) and (\ref{Eq_18}), respectively{\textemdash}using the data depicted in Fig. \ref{fig:3} and the parameters obtained from the fitting of the EIS spectra to the equivalent circuit depicted in Fig. \ref{fig:1}(b). Note that $\tilde{C}_\mathrm{F}(\omega)$ and $\tilde{G}_\mathrm{op}^\mathrm{F}(\omega)$ only contain the contributions of the Faradaic impedance; that is, they are corrected for the effect of the high-frequency impedance $\tilde{Z}_\mathrm{hf}(\omega)$. For bias potentials equal to and higher than $2.60~\mathrm{V}~\mathrm{vs}.~\mathrm{Li}/\mathrm{Li}^+$, the curves representing $\mathrm{Re}\{\tilde{G}_\mathrm{op}^\mathrm{F}(\omega)\}$ and $\mathrm{Im}\{\tilde{G}_\mathrm{op}^\mathrm{F}(\omega)\}$ were conveniently multiplied by the indicated factors so that $\mathrm{Re}\{\tilde{G}_\mathrm{op}^\mathrm{F}(\omega)\}$ and $\mathrm{Re}\{\tilde{C}_\mathrm{F}(\omega)\}$ coincided in the plots at $f=10~\mathrm{mHz}$. For the bias potentials of $2.10$ and $1.50~\mathrm{V}~\mathrm{vs}.~\mathrm{Li}/\mathrm{Li}^+$, the same scaling criterion was used but, in these cases, $\mathrm{Re}\{\tilde{G}_\mathrm{op}^\mathrm{F}(\omega)\}$ was made to coincide with $\mathrm{Re}\{\tilde{C}_\mathrm{int}(\omega)\}$ instead of $\mathrm{Re}\{\tilde{C}_\mathrm{F}(\omega)\}$. The data sets are presented from $10~\mathrm{mHz}$ up to the highest frequencies shown in the complex optical capacitances in Fig. \ref{fig:3}. The individual contributions of the complex capacitances $\tilde{C}_\mathrm{CPEad}(\omega)$ (red dashed line) and $\tilde{C}_\mathrm{int}(\omega)$ (magenta dash-dotted line){\textemdash}describing respectively the adsorption and intercalation branches of the parallel array in Fig. \ref{fig:1}(b){\textemdash}are shown in the plots. The shaded areas portray one standard deviation around the experimental values of $\mathrm{Re}\{\tilde{G}_\mathrm{op}^\mathrm{F}(\omega)\}$ and $\mathrm{Im}\{\tilde{G}_\mathrm{op}^\mathrm{F}(\omega)\}$.}
\end{figure*}

Figure \ref{fig:3} depicts the Nyquist representation of the total complex capacitance $\tilde{C}(\omega)$ and the complex optical capacitance $\tilde{G}_\mathrm{op}(\omega)$ for the $\textit{a}\mathrm{WO}_3$ film at the different bias potentials{\textemdash}the same data are shown in the Bode representation in Fig. S3 in the SI. Excluding the $1.50~\mathrm{V}~\mathrm{vs}.~\mathrm{Li}/\mathrm{Li}^+$ data, the fittings to $\tilde{C}(\omega)$, represented by the solid lines in Figs. \ref{fig:3}(a) and \ref{fig:3}(b), are excellent. Both $\tilde{C}(\omega)$ and $\tilde{G}_\mathrm{op}(\omega)$ consist mainly of a single arc-shaped response. At large, these quantities show similar qualitative behavior{\textemdash}see Figs. \ref{fig:3}(a) and \ref{fig:3}(c) and compare the arcs describing the same bias potential. Regarding $\tilde{C}(\omega)$, the size of the main arc increases with decreasing bias potential down to $2.10~\mathrm{V}~\mathrm{vs}.~\mathrm{Li}/\mathrm{Li}^+$, before becoming smaller at $1.50~\mathrm{V}~\mathrm{vs}.~\mathrm{Li}/\mathrm{Li}^+$, see Fig. \ref{fig:3}(a). The same trend is followed by the size of the main arcs related to $\tilde{G}_\mathrm{op}(\omega)$, see Fig. \ref{fig:3}(c). However, in contrast to Fig. \ref{fig:3}(a), for $\tilde{G}_\mathrm{op}(\omega)$ the $2.10$ and $2.60~\mathrm{V}~\mathrm{vs}.~\mathrm{Li}/\mathrm{Li}^+$ arcs are significantly close to each other, and the $1.50~\mathrm{V}~\mathrm{vs}.~\mathrm{Li}/\mathrm{Li}^+$ arc is the one that presents the smallest size. The high-frequency regions of $\tilde{C}(\omega)$ and $\tilde{G}_\mathrm{op}(\omega)$ reveal interesting details, see Figs. \ref{fig:3}(b) and \ref{fig:3}(d). First, for the $3.15~\mathrm{V}~\mathrm{vs}.~\mathrm{Li}/\mathrm{Li}^+$ case, there are clear signs of a small semicircle at high frequencies, which is present both in the electrical and the optical responses. Second, at bias potentials equal and lower than $2.90~\mathrm{V}~\mathrm{vs}.~\mathrm{Li}/\mathrm{Li}^+$ (being more notorious at $2.60$ and $2.10~\mathrm{V}~\mathrm{vs}.~\mathrm{Li}/\mathrm{Li}^+$), there are data points with $\mathrm{Re}\{\tilde{G}_\mathrm{op}(\omega)\}<0${\textemdash}see Fig. \ref{fig:3}(d) and the unfilled symbols in Fig. S3 in the SI. The former effect is assigned here to the adsorption process participating in the optical response. The latter is a feature that was also detected by Kim and colleagues in Ref. [\onlinecite{Kim1997}] and, in the given reference, was linked to effects related to high-frequency non-Faradaic capacitances. In the present work, we observed that the real part of the modified complex optical capacitance $\tilde{G}_\mathrm{op}^\mathrm{F}(\omega)$ did not exhibit negative values, see Fig. \ref{fig:4}. Hence, the apparent negative capacitance seen in $\tilde{G}_\mathrm{op}(\omega)$ comes from the uncompensated contribution of the voltage drop $\tilde{V}_\mathrm{hf}(\omega)$, which does not participate in the coloration.

Figure \ref{fig:4} depicts, in the Bode representation, the respective comparison between the real and imaginary parts of the Faradaic arrays $\tilde{G}_\mathrm{op}^\mathrm{F}(\omega)$, $\tilde{C}_\mathrm{F}(\omega)$, $\tilde{C}_\mathrm{int}(\omega)$, and $\tilde{C}_\mathrm{CPEad}(\omega)$ at the different bias potentials. At $3.15~\mathrm{V}~\mathrm{vs}.~\mathrm{Li}/\mathrm{Li}^+$ ($x=0.010$), see Figs. \ref{fig:4}(a) and \ref{fig:4}(b), the real as well as the imaginary parts of $\tilde{C}_\mathrm{F}(\omega)$ and $\tilde{G}_\mathrm{op}^\mathrm{F}(\omega)$ are almost perfectly proportional to each other, and small discrepancies are only observed at the beginning of the high-frequency wing. Interestingly, at this bias potential, it is clearly seen that the adsorption contribution (red dashed lines) participates in the coloration. In fact, its relative importance at high frequencies is much larger than that of the diffusion process (magenta dash-dotted lines). The latter becomes more relevant at low frequencies. The most remarkable aspect here is that the coloration follows both the adsorption and diffusion processes. The same observations also apply for the $2.90~\mathrm{V}~\mathrm{vs}.~\mathrm{Li}/\mathrm{Li}^+$ ($x=0.052$) case, see Figs. \ref{fig:4}(c) and \ref{fig:4}(d). However, there are some visible differences with respect to the $3.15~\mathrm{V}~\mathrm{vs}.~\mathrm{Li}/\mathrm{Li}^+$ data. First, the relative contribution of the adsorption process starts to decrease with respect to that of the diffusion. Second, due to the increase of the chemical diffusion coefficient (see Table \ref{table:II}), the diffusion response shifts toward higher frequencies, this is evident by looking at the peak in $\mathrm{Im}\{\tilde{C}_\mathrm{int}(\omega)\}$ shown in Fig. \ref{fig:4}(d) and comparing this with that in Fig. \ref{fig:4}(b). Third, the curves related to $\tilde{C}_\mathrm{F}(\omega)$ and $\tilde{G}_\mathrm{op}^\mathrm{F}(\omega)$ still show an excellent proportionality relation at low frequencies [see Figs. \ref{fig:4}(c) and \ref{fig:4}(d)]; however, the high-frequency discrepancies extend toward lower frequencies, see Fig. \ref{fig:4}(c). The data at $2.60~\mathrm{V}~\mathrm{vs}.~\mathrm{Li}/\mathrm{Li}^+$ ($x=0.163$), see Figs. \ref{fig:4}(e) and \ref{fig:4}(f), show a similar qualitative behavior as those at higher bias potentials. In this case, the trends enumerated for the $2.90~\mathrm{V}~\mathrm{vs}.~\mathrm{Li}/\mathrm{Li}^+$ data are even more accentuated. At $2.10~\mathrm{V}~\mathrm{vs}.~\mathrm{Li}/\mathrm{Li}^+$ ($x=0.551$), the low-frequency wing of $\mathrm{Re}\{\tilde{G}_\mathrm{op}^\mathrm{F}(\omega)\}$ presents a better proportionality relation with $\mathrm{Re}\{\tilde{C}_\mathrm{int}(\omega)\}$ than with $\mathrm{Re}\{\tilde{C}_\mathrm{F}(\omega)\}$, see Fig. \ref{fig:4}(g), and this is also the case for the respective imaginary parts, see Fig. \ref{fig:4}(h). The previous observations suggest that, at this bias potential, the contribution of the diffusion to the optical response is dominant with respect to that of the adsorption. In addition, similarly to what it is observed at higher bias potentials, the shapes of the high-frequency wings of the $\tilde{G}_\mathrm{op}^\mathrm{F}(\omega)$ spectra in Figs. \ref{fig:4}(g) and \ref{fig:4}(h) show evident discrepancies with those related to $\tilde{C}_\mathrm{F}(\omega)$, $\tilde{C}_\mathrm{int}(\omega)$, and $\tilde{C}_\mathrm{CPEad}(\omega)$. The EIS fitting for the $1.50~\mathrm{V}~\mathrm{vs}.~\mathrm{Li}/\mathrm{Li}^+$ case was not optimal, see Figs. \ref{fig:2} and \ref{fig:3}(a). Thus, we should be careful in drawing conclusions from the data presented in Figs. \ref{fig:4}(i) and \ref{fig:4}(j). Having said this, we can mention that the spectra depicted in Figs. \ref{fig:4}(i) and \ref{fig:4}(j) display a similar qualitative behavior as those at $2.10~\mathrm{V}~\mathrm{vs}.~\mathrm{Li}/\mathrm{Li}^+$.

\section{\label{sec4:level1}Discussion}

\begin{table*}[t]
\caption{\label{table:III} Linear frequencies at the maximum of the relaxation peaks of the respective imaginary parts of $\tilde{C}_\mathrm{F}(\omega)$ ($f_\mathrm{CF}^\mathrm{max}$), $\tilde{C}_\mathrm{AD1b}(\omega)=A^{-1}[i\omega \tilde{Z}_\mathrm{AD1b}(\omega)]^{-1}$ ($f_\mathrm{AD1b}^\mathrm{max}$), $\tilde{C}_\mathrm{int}(\omega)$ ($f_\mathrm{int}^\mathrm{max}$), and $\tilde{G}_\mathrm{op}^\mathrm{F}(\omega)$ ($f_\mathrm{op}^\mathrm{max}$) for the different bias potentials and the associated intercalation levels. The values were obtained graphically, and the errors are related to the spacing of the experimental frequency points around the given frequency values.}
\begin{ruledtabular}
\begin{tabular}{>{\centering\arraybackslash}m{1.8cm}>{\centering\arraybackslash}m{1.4cm}>{\centering\arraybackslash}m{1.8cm}>{\centering\arraybackslash}m{1.8cm}>{\centering\arraybackslash}m{1.8cm}>{\centering\arraybackslash}m{2.4cm}>{\centering\arraybackslash}m{2.6cm}}
Bias potential & $x=\mathrm{Li}/\mathrm{W}$ & $f_\mathrm{CF}^\mathrm{max}$ &  $f_\mathrm{AD1b}^\mathrm{max}$ &  $f_\mathrm{int}^\mathrm{max}$ &  $f_\mathrm{op}^\mathrm{max}$ \\
$(\mathrm{V}~\mathrm{vs}.~\mathrm{Li}/\mathrm{Li}^+)$ &  & ($\mathrm{Hz}$) & ($\mathrm{Hz}$) & ($\mathrm{Hz}$) & ($\mathrm{Hz}$) \\ \hline
$3.15$ & $0.010$ & $0.03\pm 0.1$ & $0.04\pm 0.02$ & $0.04\pm 0.02$ & $0.03\pm 0.01$\\ 
$2.90$ & $0.052$ & $0.16\pm 0.02$ & $0.36\pm 0.04$ & $0.18\pm 0.02$ & $0.15\pm 0.03$\\
$2.60$ & $0.163$ & $0.64\pm 0.07$ & $0.80\pm 0.09$ & $0.80\pm 0.09$ & $0.34\pm 0.06$\\
$2.10$ & $0.551$ & $0.36\pm 0.04$ & $0.45\pm 0.05$ & $0.50\pm 0.06$ & $0.25\pm 0.03$\\
$1.50$ & $2.013$ & $2.0\pm 0.2$ & $3.2\pm 0.4$ & $4.0\pm 0.5$ & $0.50\pm 0.06$\\
\end{tabular}
\end{ruledtabular}
\end{table*}

In this section we discuss the implications of the results presented above. We first make some comments on issues related to the equivalent circuit modelling of EIS data. Subsequently we make a detailed comparison of electrical and optical relaxation peaks and discuss their implication for the physical processes in electrochromic $\mathrm{WO}_3$ thin films.

The equivalent circuit in Fig. \ref{fig:1} could not be completely verified by the present measurements since the effects of some of the circuit elements were not visible in the experimental data. As mentioned before, neither $\tilde{Z}_\mathrm{CPEdl}(\omega)$ nor $R_\mathrm{ct}$ could be obtained from the present EIS data because the contribution of the double layer was presumably small and situated in a frequency range dominated by inductive effects. In order to resolve the contributions from these elements, not only should inductive effects be suppressed, but also measurements be performed at higher potentials where the contribution from the adsorption process becomes smaller. This was possible in EIS measurements on an $\textit{a}\mathrm{WO}_3$ WE performed in our previous work\cite{RojasGonzalez2020a} (using the same electrolyte and deposition conditions as those employed here) at potentials between $2.8$ and $3.7~\mathrm{V}~\mathrm{vs}.~\mathrm{Li}/\mathrm{Li}^+$. The question therefore arises whether effects of $\tilde{Z}_\mathrm{CPEdl}(\omega)$ and $R_\mathrm{ct}$ might be significant for the present data at $3.15~\mathrm{V}$. In the previous measurements,\cite{RojasGonzalez2020a} $R_\mathrm{ct}$ exhibited values of about $14$ and $3~\Omega$ at bias potentials of $3.2$ and $3.1~\mathrm{V}~\mathrm{vs}.~\mathrm{Li}/\mathrm{Li}^+$, respectively. In these cases, the respective values of $f_\mathrm{dl}$ were of about $536$ and $4106~\mathrm{Hz}${\textemdash}much higher than the upper frequency limits of the complex optical capacitance spectra shown here (about 10 Hz). Also, both at $3.2$ and $3.1~\mathrm{V}~\mathrm{vs}.~\mathrm{Li}/\mathrm{Li}^+$, $R_\mathrm{ad}$ was about two orders of magnitude higher than $R_\mathrm{ct}$. The previous EIS measurements were performed down to $2.8~\mathrm{V}~\mathrm{vs}.~\mathrm{Li}/\mathrm{Li}^+$. However, below $3.1~\mathrm{V}~\mathrm{vs}.~\mathrm{Li}/\mathrm{Li}^+$, the obtained values of $R_\mathrm{ct}$ were uncertain due to their small magnitude and relative size with respect to $R_\mathrm{ad}${\textemdash}as a reference, a $R_\mathrm{ct}$ of about $1~\Omega$ was estimated at $2.8~\mathrm{V}~\mathrm{vs}.~\mathrm{Li}/\mathrm{Li}^+$. The previous observations suggest that the effects of the double layer capacitance can indeed be neglected in the frequency region at which the optical capacitances were displayed in Figs. \ref{fig:3} and \ref{fig:4} (that is, below about $10~\mathrm{Hz}$). Provided that $R_\mathrm{ct}$ remains of the order of $1~\Omega$ even at lower bias potentials, it could be of a similar order of magnitude as $R_\mathrm{ad}$ at $2.1$ and $1.5~\mathrm{V}~\mathrm{vs}.~\mathrm{Li}/\mathrm{Li}^+$ (see Table \ref{table:I}). This could introduce errors in the obtention of the different elements at the mentioned bias potentials when the equivalent circuit in Fig. \ref{fig:1}(b) is used.

It should also be mentioned that a high-frequency semicircle in EIS data has frequently been assigned in previous works to the parallel combination $R_\mathrm{ct}\tilde{Z}_\mathrm{CPEdl}(\omega)$ (see for example Ref. [\onlinecite{Malmgren2017}]), without considering the adsorption process. Hence, the usual physical interpretation of high-frequency effects in EIS data needs to be re-evaluated because of the implications drawn from our present analysis.

We next discuss the relaxation peaks in the complex Faradaic electrical and optical capacitances as well as their relation to the ion diffusion and adsorption contributions to EIS data. The relaxation peaks are visible as maxima of the imaginary parts of $\tilde{C}_\mathrm{F}(\omega)$ and $\tilde{G}_\mathrm{op}^\mathrm{F}(\omega)$ in Fig. \ref{fig:4}. We compare the imaginary parts of $\tilde{C}_\mathrm{F}(\omega)$, $\tilde{C}_\mathrm{AD1b}(\omega)=A^{-1}[i\omega\tilde{Z}_\mathrm{AD1b}(\omega)]^{-1}$, $\tilde{C}_\mathrm{int}(\omega)$, and $\tilde{G}_\mathrm{op}^\mathrm{F}(\omega)$, and denote the linear frequencies at the maximum of the respective relaxation peaks by $f_\mathrm{CF}^\mathrm{max}$, $f_\mathrm{AD1b}^\mathrm{max}$, $f_\mathrm{int}^\mathrm{max}$, and $f_\mathrm{op}^\mathrm{max}$. Their values are presented in Table \ref{table:III} for the different bias potentials (together with their associated intercalation levels). Note that $f_\mathrm{AD1b}^\mathrm{max}$ is not equal to $f_\mathrm{D}$ due to the mathematical form of $\tilde{Z}_\mathrm{AD1b}(\omega)$ in Eq. (\ref{Eq_6}). In general, the former is higher than the latter. All the linear frequencies displayed a similar qualitative behavior. That is, the lowest value was located at $3.15~\mathrm{V}~\mathrm{vs}.~\mathrm{Li}/\mathrm{Li}^+$, a subsequent increase was observed with decreasing bias potential down to $2.60~\mathrm{V}~\mathrm{vs}.~\mathrm{Li}/\mathrm{Li}^+$, where a local maximum could be distinguished. Thereafter, the linear frequencies decreased at $2.10~\mathrm{V}~\mathrm{vs}.~\mathrm{Li}/\mathrm{Li}^+$ and increased again at $1.50~\mathrm{V}~\mathrm{vs}.~\mathrm{Li}/\mathrm{Li}^+${\textemdash}which, in all cases, corresponded to the maximum value. This kind of behavior is consistent with that of the ion-electron couple dynamics described by the chemical diffusion response in Table \ref{table:II}{\textemdash}in particular, in relation to $f_\mathrm{D}$ and $D_\mathrm{ch}$. The values of $f_\mathrm{AD1b}^\mathrm{max}$ and $f_\mathrm{int}^\mathrm{max}$ are similar for all the bias potentials, with the highest relative difference at $2.60~\mathrm{V}~\mathrm{vs}.~\mathrm{Li}/\mathrm{Li}^+$. The discrepancies are due to the presence of $R_\mathrm{ad}$ in the so-called intercalation branch. The complex Faradaic capacitance $\tilde{C}_\mathrm{F}(\omega)$ accounts for both the contributions of the adsorption and the intercalation branches, which are given by $\tilde{C}_\mathrm{CPEad}(\omega)$ and $\tilde{C}_\mathrm{int}(\omega)$, respectively. As shown in Table \ref{table:III}, $f_\mathrm{int}^\mathrm{max}$ is higher than $f_\mathrm{CF}^\mathrm{\max}$ at the given bias potentials. Thus, the adsorption branch generates a delay in the relaxation of $\tilde{C}_\mathrm{F}(\omega)$ with respect to that of $\tilde{C}_\mathrm{int}(\omega)$. In general, the values of  $f_\mathrm{op}^\mathrm{max}$ are similar or lower than $f_\mathrm{CF}^\mathrm{max}$, and their difference tends to increase with decreasing bias potential. However, a decrease in their difference is observed at $2.10~\mathrm{V}~\mathrm{vs}.~\mathrm{Li}/\mathrm{Li}^+$ with respect to the neighboring bias potentials. Interestingly, at this bias potential, the same feature (that is, a local minimum) is observed in the overall behavior of the linear frequencies as well as in the chemical diffusion coefficient. Actually, the relaxation of the modified complex optical capacitance $\tilde{G}_\mathrm{op}^\mathrm{F}(\omega)$ is almost perfectly synchronized with that of $\tilde{C}_\mathrm{F}(\omega)$ at high bias potentials (that is, $3.15$ and $2.90~\mathrm{V}~\mathrm{vs}.~\mathrm{Li}/\mathrm{Li}^+$), as depicted in Figs. \ref{fig:4}(b) and \ref{fig:4}(d) as well as in Table \ref{table:III}. These cases correspond to situations in which the adsorption branch gives an appreciable contribution to the coloration. On the other hand, a noticeable delay in the relaxation of $\tilde{G}_\mathrm{op}^\mathrm{F}(\omega)$ with respect to that of $\tilde{C}_\mathrm{F}(\omega)$ can be seen at low bias potentials. In these cases, the coloration response qualitatively seems to resemble a delayed response, similar in shape to $\tilde{C}_\mathrm{int}(\omega)$.

Currently, there is no conclusive explanation for the previously commented points. However, we can hypothesize the following. In order for the coloration to take place, the ion-electron couple not only has to find an empty site to occupy, it also needs to find one with empty neighboring sites to make possible the transitions that are responsible for the coloration. This is more difficult at high intercalation levels and would lead to a delay in the optical response. Indeed, finding a site with an empty neighbor would be relatively easy at low intercalation levels, such as in the $3.15$ and $2.90~\mathrm{V}~\mathrm{vs}.~\mathrm{Li}/\mathrm{Li}^+$ cases, where there is a good agreement between $\tilde{G}_\mathrm{op}^\mathrm{F}(\omega)$ and $\tilde{C}_\mathrm{F}(\omega)$. On the other hand, at high intercalation levels, the surface (where the adsorption effects are relevant) may be almost saturated, and the ion-electron couple would have to diffuse further into the film in order to find appropriate sites for the coloration. This may explain why, at lower bias potentials, the impact of the adsorption effects on the coloration is smaller. According to the previous arguments, it is also reasonable to expect differences between electrical and optical responses, and especially it appears probable that the optical response would be delayed at high intercalation levels. Thus, it may be sensible to use different sets of effective diffusion parameters to describe the electrical and optical spectra, as done for example by Agrisuelas et al. in Ref. [\onlinecite{Agrisuelas2009a}].

Another interesting observation related to Fig. \ref{fig:4} concerns the behavior of the multiplying factor needed to bring the optical spectra into coincidence with the electrical one. It is seen that the lower the bias potential the higher the numerical factor that multiplies the modified complex optical capacitance. This can be explained in terms of the differential coloration efficiency, which is defined as the ratio between the complex optical and the electrical capacitances.\cite{Rojas-Gonzalez2019} Indeed, in a previous work,\cite{RojasGonzalez2020} it was shown that the amplitude of the differential coloration efficiency of $\textit{a}\mathrm{WO}_3$ (experimentally) decreases with decreasing bias potential (see figure 4 in Ref. [\onlinecite{RojasGonzalez2020}]).

Finally, we discuss some effects that may influence the optical capacitance at medium and high intercalation levels{\textemdash}that is, at low applied potentials. It has been suggested that the different electronic transitions related to the coloration{\textemdash}that is, $\mathrm{W}^{5+}\rightarrow\mathrm{W}^{6+}$, $\mathrm{W}^{4+}\rightarrow\mathrm{W}^{5+}$, and $\mathrm{W}^{4+}\rightarrow\mathrm{W}^{6+}${\textemdash}give rise to significantly dissimilar optical absorption spectra.\cite{Berggren2007} This could be one of the reasons for the observed discrepancies between the complex optical and electrical capacitances in Fig. \ref{fig:4}.  The presence of $\mathrm{W}^{4+}$ sites would make the transitions of the electrons to exhibit three different absorption processes, each of them with its own absorption strength depending on the photon energy (or optical wavelength). Hence, by performing SECIS measurements at different optical wavelengths, it would be possible to ascertain the effect of the $\mathrm{W}^{4+}$ sites on the optical capacitance. This is out of the scope of this work but is certainly interesting for future studies. SECIS measurements on $\textit{a}\mathrm{WO}_3$ performed at optical wavelengths of $470$, $530$, and $810~\mathrm{nm}$ and at a bias potential of $2.6~\mathrm{V}~\mathrm{vs}.~\mathrm{Li}/\mathrm{Li}^+$ were presented elsewhere.\cite{Rojas-Gonzalez2019} They showed that the related complex optical capacitance spectra were, within experimental error, perfectly proportional to each other. This could be an argument against ascribing the mentioned discrepancies noticed in Fig. \ref{fig:4} to the $\mathrm{W}^{4+}$ sites. However, $2.6~\mathrm{V}~\mathrm{vs}.~\mathrm{Li}/\mathrm{Li}^+$ corresponds to $x\approx 0.163$ and, at this intercalation level, the expected probability of a transition involving $\mathrm{W}^{4+}$ sites is still much smaller than that of $\mathrm{W}^{5+}\rightarrow\mathrm{W}^{6+}$ (see figure 1 in Ref. [\onlinecite{Berggren2007}]). Thus, this is not a conclusive result and more studies at higher intercalation levels are required in this regard.

At potentials below about $2~\mathrm{V}~\mathrm{vs}.~\mathrm{Li}/\mathrm{Li}^+$, there are other effects that may influence dynamic measurements like EIS and SECIS. At these high intercalation levels, the electrochromism of $\mathrm{WO}_3$ thin films degrades during long-time cycling between bleached and colored states due to irreversible trapping of $\mathrm{Li}$ ions.\cite{Arvizu2015,Baloukas2017} The same effect would be seen if a sample is held at low bias potentials for a long time. At these low potentials, a number of $\mathrm{Li}$ compounds can form\cite{Bressers1998} and this is a possible reason for the trapping of $\mathrm{Li}$. In our previous work,\cite{RojasGonzalez2020} we documented an underlying bleaching process during SECIS measurements below $2~\mathrm{V}~\mathrm{vs}.~\mathrm{Li}/\mathrm{Li}^+$, and this might be due to the formation of non-absorbing $\mathrm{Li}$ compounds. However, it is not known to which extent, and if so how, these relatively slow processes would influence the EIS and CIS spectra.

\section{\label{sec5:level1}Conclusions}

The electrochromic response of an amorphous tungsten oxide thin film was studied by simultaneous frequency-dependent electrical and optical measurements at different intercalation levels. A comparison between the complex electrical and optical capacitances gives valuable information about the coloration mechanisms in electrochromic systems. The following was observed in the case of amorphous tungsten oxide. Two processes are responsible for the coloration{\textemdash}that is, superficial adsorption effects and the diffusion of the electron-ion couple in the film. At low intercalation levels, the electrical and optical responses were nearly identical (apart from a multiplying constant), and both the adsorption as well as the diffusion processes contributed considerably to the coloration. On the other hand, at high intercalation levels, clear discrepancies between the electrical and the optical responses were noticed, and the coloration mostly followed the diffusion process (in comparison with the apparent contribution of the adsorption effects), although the optical response appeared to be significantly delayed. This delay may be related to the difficulty of finding a site with an empty neighboring one at high intercalation levels. The use of different optical wavelengths in frequency-dependent measurements (like those presented in this work) could shed light on diverse electrochromic-related phenomena{\textemdash}for example, the influence of $\mathrm{W}^{4+}$ sites on the observed discrepancies between the optical and electrical responses in amorphous tungsten oxide. Furthermore, the possibility of distinguishing between diffusion- and adsorption-related effects is of particular interest for electrochromic systems in which the coloration has previously been mainly assigned to the former.

\begin{acknowledgments}
This work was supported by a grant from the Swedish Research Council (No. VR-2016-03713). E. A. Rojas-Gonz{\'a}lez is grateful for the scholarship granted by the University of Costa Rica for pursuing doctoral studies at Uppsala University. Daniel Primetzhofer, Marcos Moro, and the staff at the Tandem Laboratory at Uppsala University are thanked for assistance with RBS measurements. Support by VR-RFI (contract 2017-00646-9) and the Swedish Foundation for Strategic Research (contract RIF14-0053) for accelerator operation is gratefully acknowledged.
\end{acknowledgments}

\section*{Data Availability}
The data that support the findings of this study are available from the corresponding author
upon reasonable request.


%

\end{document}